\documentstyle[twoside,fleqn,espcrc2,epsf]{article}
\let\useblackboard=\iftrue
\let\includefigures=\iftrue
\let\includefigures=\iffalse
\newcommand{\eqa}{\begin{eqnarray}}
\newcommand{\ena}{\end{eqnarray}}
\newcommand{\eqar}{\begin{array}}
\newcommand{\enar}{\end{array}}
\newcommand{\eqn}{\begin{equation}}
\newcommand{\enq}{\end{equation}}
\newcommand{\bl}{\begin{enumerate}}
\newcommand{\el}{\end{enumerate}}

\useblackboard
\font\blackboard=msbm10 scaled \magstep1
\font\blackboards=msbm7
\font\blackboardss=msbm5
\newfam\black
\textfont\black=\blackboard
\scriptfont\black=\blackboards
\scriptscriptfont\black=\blackboardss
\def\Bbb#1{{\fam\black\relax#1}}
\else
\def\Bbb{\bf}
\fi
\def\yboxit#1#2{\vbox{\hrule height #1 \hbox{\vrule width #1
\vbox{#2}\vrule width #1 }\hrule height #1 }}
\def\fillbox#1{\hbox to #1{\vbox to #1{\vfil}\hfil}}
\def\ybox{{\lower 1.3pt \yboxit{0.4pt}{\fillbox{8pt}}\hskip-0.2pt}}
\def\comments#1{}

\def\BR{\Bbb{R}}
\def\BZ{\Bbb{Z}}
\def\p{\partial}
\def\ap{\alpha'}

\def\bZ{{\bar Z}}

\def\bi{{\bar i}}

\def\STr{{\rm Str\ }}
\def\Tr{{\rm tr\ }}
\def\tr{{\rm tr\ }}

\def\CM{{\cal M}}

\def\CL{{\cal L}}

\def\CN{{\cal N}}

\def\ALE{{\rm ALE}}
\def\K3{{\rm K3}}
\title{\vbox{\hbox{\rightline{\rm\small RU-97-66}}
\hbox{D-branes and Matrix Theory in Curved Space}}}
\author{Michael R. Douglas\thanks{
Address from Sept. 1--Dec. 31, 1997:
Institut des Hautes Etudes Scientifiques,
Bures-sur-Yvette, France.}\\
Dept. of Physics and Astronomy\\
Rutgers University\\
Piscataway, NJ 08855
        }
\begin{document}

\begin{abstract}
We discuss the relation between supersymmetric gauge theory of branes
and supergravity; as it was discovered in D-brane physics, and as it
appears in Matrix theory, with emphasis on motion in curved
backgrounds.
We argue that gauged sigma model
Lagrangians can be used
as definitions of Matrix theory in curved space.

Lecture given at Strings '97; June 20, 1997.
\end{abstract}

\maketitle

\section{Introduction}

In this lecture, we will discuss a class of quantum mechanical
actions which we believe
are one appropriate starting point for defining Matrix theory
in curved backgrounds.
This is in part based on the works
\cite{dcurve} and \cite{DOS},
and on work in progress with A. Kato and H. Ooguri.
But first, let us give some historical background.

A central lesson from the physics of D-branes
\cite{DLP,Polchinski,Polrev}
was a new and potentially deep relationship between
supersymmetric gauge theory and supergravity.
The prototype for this was
the computation of the force between two D-branes in
\cite{Polchinski,Bachas}.
At leading order in the string coupling, this is found
by evaluating the world-sheet path integral on an annulus with
one boundary on each D-brane.
Although the static force between parallel branes vanishes,
by considering velocity-dependent forces or by turning on fields
on the branes, one finds non-zero interactions, which can be
understood in field theory terms in two ways: either as the
sum of all classical
closed string exchanges between the branes, or using
world-sheet duality, as the sum of one-loop amplitudes in the
gauge theory of the open strings ending on the branes:
\eqa\label{sumexp}
\sum_{{\rm closed\atop string mode} i} v^n Q_1 Q_2 G(x,y; m_i)
= \\
\sum_{{\rm open\atop string mode} i} v^n \int {d^{p+1}k\over (k^2+m_i^2)^n}
\ena
with $v$ velocity,
$Q_i$ appropriate charges and $G$ the Green function.
The annulus amplitude is quite accessible to explicit computation,
and was explored in
\cite{Li,Lifschytz}
and many other works.

In general this is a relation between two descriptions within
string theory, and requires the sum over the entire string spectrum
for its validity, just like the modular invariance of closed string
amplitudes.
However, it was soon discovered that in certain amplitudes with
residual supersymmetry (left unbroken by the velocities or
field expectation values), the massive string states decouple,
and the relation becomes a relation between
interactions computed in two field theories: supergravity, and
the gauge theory of the lightest open strings stretched between
the two branes: maximally supersymmetric Yang-Mills theory in
the case of parallel branes in flat space, and more generally
an SYM with matter determined by the brane configuration.

The first example of this phenomenon (to our knowledge)
was found in \cite{DouglasLi}.  After obtaining $d=4$, $\CN=4$ SYM
from parallel $3$-branes, it was natural to ask what D-branes
had to say about $\CN=2$ SYM, and in that work pure $\CN=2$ SYM was
obtained by the simple expedient of wrapping $7$-branes on K3.
It was found that the one-loop prepotential in the gauge theory
(which reduces to a sum over BPS states) was equal to the classical
Green function for exchange of massless fields in the supergravity.

One interesting followup (for the issues raised in this talk)
to this was \cite{BachasFabre}, where it was shown that such gauge theory
computations are finite, despite the lack of any explicit
UV cutoff, thanks to cancellations between the loop divergences.
This is `dual' to the statement that, even in
two dimensions, the Greens
function (which determines the supergravity amplitude)
will not have an IR divergence on a compact space.
The result of \cite{DouglasLi}
was also what led us to look for and find the analogous
relation
for the $v^4/r^7$ interaction between D$0$-branes
in \cite{DKPS}, and thus the possibility of computing this
interaction in gauge theory.

These observations found a natural place, along with other connections
between supersymmetric quantum mechanics and supergravity
(most notably, the description of the supermembrane developed in
\cite{DHN} and first cited in this context in \cite{Townsend}),
and many other observations in D-brane physics,
as part of the far-reaching Matrix theory conjecture of
Banks, Fischler, Shenker and Susskind \cite{BFSS}.
To put this in a nutshell, all of eleven-dimensional physics
(or what is visible in the infinite momentum frame) is contained
in maximally supersymmetric gauge theory, reduced to quantum
mechanics and in the large $N$ limit.
The result of \cite{DKPS} then explains the leading long-distance
supergravity interaction between D$0$-branes --
it is produced as a one-loop effect in the quantum mechanics.

These observations
also play an important role at weak string coupling --
but there, they are a special case of a different relation \cite{DKPS}:
gauge theory replaces gravity for D-branes at substringy distances
and low velocities, but can in general give different predictions.

Let us compare the two limits in the context of D-branes in a
background with spatial curvature.  Both are
potentially relevant to low-energy physics.
After the comparison, we will concentrate on the large $R_{11}$
(strong string coupling) limit,
which after all is a new and fascinating regime which has become
accessible to us by virtue of \cite{BFSS}, but it is helpful to have
the larger picture in mind.

\subsection{Weak string coupling -- $l_s >> l_p$}

\bl
\item
The D-brane world-volume action is defined by world-sheet computations in
superstring theory, along the lines of \cite{DLP,Leigh},
so can in principle be computed in any background.

\item
At substringy distances, $r\le l_s$, gauge theory replaces gravity.
The fact that gravity is produced by integrating out stretched
strings, and the relation between their mass and separation
$m = T_s r$, implies that the UV limit for gravitational interactions
is defined by the IR physics of the branes.

\item
Even if the string coupling at
infinity is weak, quantum effects on the branes can be
enhanced by IR effects.  In field theory language, the couplings
can grow under renormalization.
For D$0$-branes, loop effects are controlled
by the dimensionless parameter $g_s(\ap)^{3/2} /r^3 \sim (l_{p11}/r)^3$.

\item
A convenient way to study this type of `gravity' is to introduce an
auxiliary D-brane `probe' on which open strings can end,
and solve its gauge theory.
We can interpret its moduli space,
or more generally the configuration space visible at low energy,
as the space-time geometry.
This allows bringing all the techniques
of supersymmetric gauge theory to bear, and thus can provide exact
results for the metric and other fields.

\item
For $r > l_p$, these results need not agree
with the predictions
of ten or eleven-dimensional supergravity.
In special cases, supersymmetry constrains the Lagrangian to force
such agreement, but in general, there is a non-trivial interpolation
between the long distance and short distance behavior, with a cross-over
at the string scale.

\item
At long distances $r > l_s$, gravity replaces gauge theory --
the interaction is better thought of as a sum over closed string
states, which at low energies reduces to supergravity.
Normally one thinks of the infinite sum over open string states
as regulating the open string theory, but we could phrase this
relation in a different way:
the UV limit of the gauge theories on the branes is defined by
the IR behavior of supergravity.

\el

Let us return to point 5, and the general statement that D-branes
see both supergravity and gauge theory in different limits.
This is how exact results for the annulus
diagram generically behave.  An interesting example can be found
in the system of a D$0$-brane and a D$6$-brane.
Although this
breaks supersymmetry completely, it does so in a controllable way --
the leading interaction is a repulsive potential, with two different
limiting behaviors.
The supergravity interpretation of the D$6$-brane is a KK monopole,
around which the D$0$-brane sees a $1/r$ potential, while at short distances
the potential is produced by integrating out fermionic stretched strings
and has the generic quantum mechanical behavior $V \sim -r$.
In general, a non-constant potential in the probe theory corresponds to
a non-constant $g_{00}$ component of the probe metric, and we conclude
that the D$0$-brane does not see the KK monopole metric at all
scales.\footnote{It is interesting that by turning on gauge fields on
the 6-brane, one can get the $1/r$ behavior at all scales
\cite{Lifschytz46}.}

This system is also a good example of the phenomenon (noted in
a different context in \cite{BanksSeiSil})
of different probes seeing different
metrics, as the D$2$ brane will see the KK monopole metric at all distances.
In string theory, this is no contradiction as different probes can have
different couplings to the massive closed string states, which from the
world-volume point of view also contribute to the metric.
Which probe sees ``the'' metric?
In the example at hand, it is the D$2$-brane,
which preserves enough supersymmetry to forbid such couplings, but in
general there is no such argument and one would say only that the metric
seen by the lightest objects is the most relevant one physically.

When does supersymmetry determine the metric ?
The essential distinction is
between backgrounds breaking half the supersymmetry (e.g. ALE
spaces or K3), and those breaking more.  Eight real supersymmetries
guarantee that the target space is hyperk\"ahler (assuming the
non-metric fields are zero) and thus that it satisfies the equation
of motion.  Four real supersymmetries are only enough to guarantee
that it is K\"ahler, and do not imply specific equations of motion.

In fundamental string theory, the sigma model metric does not in
general satisfy the low-energy supergravity equations of motion.
These receive corrections \cite{Grisaru,Gross}
\eqn\label{betafn}
0 = \beta^{(g)}_{\mu\nu} = R_{\mu\nu} + \ap^3 R^4 + \ldots
\enq
The sigma model metric is not directly observable and suffers
from renormalization prescription ambiguities, but the D$0$-brane
metric is observable.
It is defined by a similar calculation, which at
this writing has not been done, but has no reason not to also receive
corrections.
As Greene described in his lecture here,
there is an alternate (D-brane on orbifold) technique
for getting at this metric, and the results from this also suggest that
it will not be Ricci flat. \cite{DG}

\subsection{Large $R_{11}$ as defined by Matrix theory}

I will be brief, as this has been discussed by many speakers
here.  Furthermore, some of the points (2, 3 and 4) are the same
in both contexts.  However we have

\begin{list}{$\foo'$.}{}
\let\foo=1
\item
At present we can only conjecture that some action
(or ``base theory'') appropriate
for each background exists (more on this in the next section).
In general, this may not be the same as any
weak coupling D-brane action.

An explicit example of this was given in \cite{DOS}.
At weak string coupling, the properties of D$0$-branes moving
on the K3 manifold are explicitly calculable in the orbifold
limit and fairly well understood in general, and using this
it was shown that the annulus amplitude could not reproduce
the supergravity interaction on K3 without including all excited
open string states, and relying on world-sheet duality to relate
this to closed string exchange.

What this means is not that Matrix theory cannot be defined on K3,
only that keeping only the lightest states in the D$0$-brane
action derived at weak string coupling is not a correct definition.
Even if the correct action involves the same degrees of freedom,
terms not protected by supersymmetry
can receive arbitrary corrections in the large $R_{11}$ limit.
The non-renormalization arguments of \cite{BFSS} do not generalize
to reduced supersymmetry.

\let\foo=5
\item
For $r > l_{p11}$, gauge theory results should agree with
predictions of supergravity in the IMF.
In particular, on a background with small curvature $R l_{p11}^2 <<
1$, we should have
\bl
\item
The background must satisfy the equation of motion
\eqn\label{einstein}
R_{\mu\nu} = 0
\enq
\item
The leading interactions between gravitons are those of linearized
gravity,
\eqa\label{gravity}
H_{eff} &= v^4 G(x,y) \\
G(x,y) &{\sim\atop{d\rightarrow 0}} {1\over d^7(x,y)} + {R\over d^5} +
 {R^2\over d^3} + \ldots \label{eq:green}
\ena
where $d=d(x,y)$ is the distance (measured along the shortest
geodesic) between the locations $x$ and $y$ of the branes.
The short distance expansion for the Green function $G(x,y)$ can
be derived using heat kernel techniques \cite{deWitt}.
\el

It may be necessary to take the large $N$ limit to get this
agreement, as stated in the original proposal.
To put this more physically, the elementary states are not just the
D$0$-branes of the weak coupling limit, but bound states involving
arbitrary numbers of D$0$-branes, and it could be that the structure
of the bound states is important in the dynamics.

More recently, Susskind has proposed that the finite $N$ gauge
theories also have an M theory interpretation, as the theory with
a compact light-like dimension (sometimes called DLCQ) \cite{Suss-finite}.
This leads to very strong predictions which have found quite non-trivial
support, as described by the Beckers here.

\let\foo=6
\item
Gauge theory must reproduce gravity at all scales $r > l_p$ with
no upper limit.
Now the IR behavior of gravity corresponds to UV behavior in the gauge
theory, because it is determined by effects from integrating out
super-massive states.
Therefore, it would seem, this UV behavior must be well-defined.
\end{list}

As we will review shortly, compactification of Matrix theory
involves not just quantum mechanics but quantum field theory,
and point ($6'$)
appears to predict that field theory has far more possible
UV limits than we ever dreamed; indeed, one for each possible IR
behavior in supergravity.

I would like to advocate a point of view closer to (6) above --
eventually, we will come to regard the UV behavior of these theories
as defined by the IR behavior of the corresponding gravity.
The relation between gauge theory and gravity will be much more symmetric
than in the Matrix theory work so far.

\medskip

This comparison was intended to emphasize the similarities as well
as the differences between the two frameworks.  Ultimately, if
Matrix theory provides a complete formulation,
it should
be possible to derive all weak string coupling results from it,
perhaps along the lines of \cite{DVV}.  In their picture,
achieving the weak string coupling limit requires a non-trivial RG flow,
which is consistent with point ($1'$) made above.

\section{Matrix theory in curved space}

The original conjecture of \cite{BFSS} did not claim to define M
theory in a general curved background.
The curved backgrounds which it does treat are
those which can be realized by inserting objects formed
out of zero-branes as blocks in the large $N$ matrix, as described
here by Dijkgraaf.
This can clearly realize time-dependent backgrounds
such as propagating gravitational waves.
It can also realize some time-independent backgrounds, such as the
longitudinal five-brane \cite{BSS}.

It is an interesting question to what extent the original conjecture
covers the most general curved background.  At present, there is
no evidence that time-independent
backgrounds with non-trivial metric but zero
three-form tensor, or with non-trivial topology can be described.

On the other hand, one might generalize the original conjecture
and propose new ``base theories'' which serve as the
definition of Matrix theory on curved backgrounds.
Indeed, the general wisdom about field theory in
the infinite momentum frame is that a change in the vacuum must be
represented as a change in the light-cone Hamiltonian.
An example of this was the proposal in \cite{BerkoozDouglas}
to represent the longitudinal five-brane by adding a hypermultiplet
to the gauge theory.

The most straightforward way to put the theory in curved space is
the following.
A single D-brane in curved space will be described by a
Nambu-Born-Infeld action, which in the $\alpha'\rightarrow 0$ limit
reduces to decoupled super-Maxwell and nonlinear sigma model actions.
The logical generalization of this to $N$ D-branes is to promote
the sigma model coordinates to matrices, and
define Matrix theory using
some supersymmetrized version of the action
\eqa\label{eq:sigma}
\int dt\ &\tr g_{ij}(X) D_t X^i D_t X^j +  \\
&\tr g_{ij}(X) g_{kl}(X) [X^i,X^k][X^j,X^l] . \nonumber
\ena

However, this expression is highly ambiguous, as we must now choose
an ordering prescription for the matrices $X^\mu$.
Furthermore, since
quantum mechanics has no obvious analog of the renormalizability
constraint on quantum field theory, there are an infinite number
of higher derivative terms we might add, and reduced supersymmetry
gives only weak constraints on these.

What we will shortly propose, is that these ambiguities will be
resolved by requiring that the IR gravitational physics be correctly
reproduced, and show how this could work for the two points in (3') above.

Before we do this, let us briefly mention some of the interesting
new elements which appear when space has non-trivial topology.
Non-trivial homology leads to new conserved charges
and new BPS states, such as the membrane and five-brane wrapped
around the homology cycle.
An attractive feature of Matrix theory is that these are just
as fundamental as the original D$0$-branes.
The new conserved charges correspond to topological charges in
the gauge theory \cite{Banks}.

In \cite{Taylor,BFSS},
a very simple class of modified Hamiltonians was proposed
to describe toroidal compactification: compactifying $p$ dimensions
is accomplished by replacing D$0$-branes with D$p$-branes.
%
The prescription is justified by constructing the torus
as a quotient $\BR^p/\BZ^p$, where $\BZ^p$ acts both on space-time
and on the gauge indices: for each vector $e_I$ in the $\BZ_n$ lattice
we have
\eqa\label{eq:torquot}
X^i + e_I = U^\dag_I X^i U_I .
\ena

Now, two new BPS states which arise are the KK states and wrapped
strings, which correspond respectively to the electric and magnetic
fluxes of the gauge theory.  This leads to the beautiful result that
T-duality, which only becomes a symmetry after compactifying three
dimensions, is exactly S-duality of the underlying $3+1$ gauge theory!
\cite{GRT,Suss-dual}

A similar orbifold prescription can be used to define
orbifolds in the traditional string theory sense such as $T^4/\BZ_2$
\cite{FR} or $T^6/\BZ_3$.
Now an interesting part of this physics is localized to the fixed
points, and one can study this in the context of the simpler orbifolds
$\BR^4/\BZ_2$ or $\BR^6/\BZ_3$ \cite{DM,egs,DOS,DGM}.
As explained here by Greene, these are linear sigma models with
Fayet-Iliopoulos terms, whose moduli spaces are smooth
ALE spaces asymptotic to the original orbifold.

These models contain more degrees of freedom than the non-linear sigma
model (\ref{eq:sigma}), and there is a good physical reason for this.
Orbifolds are  singular limits, but the physics must remain
non-singular in this limit.  This is possible because the additional
quantum mechanical degrees of freedom become massless in the limit.

All of these ALE spaces have non-trivial two-cycles and thus
these theories also contain wrapped membranes.  As argued in
\cite{PolPro,egs,DGM}, these are also distinct ``topological'' sectors
of the quantum mechanics, realized by modifying the orbifold construction
to use general representations of the point group (intuitively, leaving
out some of the images of the D$0$-branes, which produces objects bound
to the fixed point).

As described here by Seiberg, the gauge theory prescription
encounters difficulties on compactifying more dimensions.  
This has
led to a fascinating series of works in which a series of six-dimensional
string theories have been used to compactify Matrix theory
on $T^4$ and $T^5$.  A simple argument for the role of string theory on
$T^5$ is that the M theory duality group $SO(5,5;\BZ)$ can be directly
identified with the T-duality group of a string theory on $T^5$.
\cite{DVVfive}

Using these theories, 
Govindarajan \cite{Gov}\ and 
Berkooz and Rozali \cite{BerkoozRozali}\ have proposed to define Matrix
theory on $\K3\times S^1$ as the six-dimensional string theory
compactified on the
dual space, again topologically $\K3\times S^1$. 
As Berkooz described
here, this reproduces the appropriate string dualities, notably to the
heterotic string theory, and leads to a simple origin for the additional
degrees of freedom in the existing Matrix constructions of the heterotic
string.

Let us add to their evidence the comment \cite{DOS}
that requiring that the model
be non-singular in the orbifold limit also appears to favor string
theory over proposals using gauge theory such as \cite{FR}.
Field theories are typically singular in the orbifold limit, and this
appears to be the case for $4+1$ gauge theory on $T^4/\BZ_2$. \cite{DougFinn}
On the other hand,
it seems reasonable to hope that the good behavior of string theory
on orbifolds will carry over to the six-dimensional string.

\section{Gravity from gauge theory}

As yet, none of the proposals mentioned in the previous
section have passed the two tests
that the moduli space should be a symmetric
product of Ricci-flat metrics, and that the supergravity interaction
should be correctly reproduced.
Testing the supergravity interaction is not easy and so we should
try to do this in the simplest context possible.

Given target space locality, the simplest models to test will
be those with non-compact target space, for several reasons.
On the practical side, these metrics and Green functions are much
simpler.  There is little hope to explicitly write the Green function
on K3; even the metric is not known.
Conceptually, in order to make the first test, we need to study
backgrounds which are not solutions as well as those which are;
the meaning of the ``dual manifold'' used in the compactification
constructions is not at all clear in this case.

One can argue with the assumption of target space locality --
indeed, Banks emphasized the non-locality of the theory in his talk here.
We will discuss this point at length in section 6.

Even the proposed definition on $\BR^7\times \ALE$ studied in
\cite{DOS} is more complicated
than we want, because of the non-trivial topology.
The simplest model to consider is clearly (\ref{eq:sigma}).

Thus we seek $U(N)$ gauged non-linear sigma models with a specified
metric, and which reproduce the supergravity interaction
as a one-loop effect.
As we said, the first issue in using (\ref{eq:sigma}) is to resolve
the matrix ordering ambiguities.
Now for the problem at hand, only a small part of this ambiguity
will be
important, because we are only going to consider linearized fluctuations
around the moduli space (which will again be diagonal matrices)
to compute our one-loop amplitude.  These will only see terms which
are up to second order in commutators $[X^\mu,X^\nu]$.

How will we reproduce the supergravity interaction?
To get the leading $v^4/d^7(x,y)$ term at short distances,
we need a gauge theory in which the $U(N)$ gauge action is the
same as in flat space, $X^i \rightarrow U^\dag X^i U$,
but in which all states
which had mass $m \propto r$ in flat space now have mass $m \propto
d(x,y)$.
More explicitly, whatever form of (\ref{eq:sigma}) we take,
one loop amplitudes will only depend on the expansion of the
Lagrangian to quadratic order in the off-diagonal matrix elements
$W$ and $\theta$.
The mass condition requires this to take the form
\eqa\label{eq:one-loop}
\CL_{od} = &K_B (DW)^2 - K_B d^2(x,y) W^2 +  \\
&iK_F \theta D \theta + K_F \theta \Gamma_i m_F^i \theta \nonumber,
\ena
where $K_B$ and $K_F$ are arbitrary functions of the curved space
positions,
and $\Gamma_i m_F^i$ is a matrix with eigenvalues $\pm d(x,y)$.

If the velocity $v$ and polarizations are purely in the flat directions,
we can remove $K_B$ and $K_F$ by rescaling the fields.
Then, since the gauge coupling is universal,
the one-loop gauge theory computation of \cite{DKPS,BFSS}
proceeds in exactly the same way,
and enjoys the same supersymmetric cancellations, with the only difference
being the replacement $m \rightarrow d(x,y)$.

A similar computation can be done for velocity $v$ or polarizations
in the curved dimensions, and getting this right requires additional
conditions relating the boson and fermion kinetic terms, which remain
to be formulated precisely.

The discussion so far allows us to formulate necessary conditions
for our model to pass the two tests.  Indeed, the
condition on the mass of the stretched strings is intuitively obvious;
the only surprise is that this is not automatic in the gauge theory
description.

\section{D-geometry}

Let us state the problem in a self-contained way which we could
give to a mathematician:
Given a $d$-dimensional manifold with metric $\CM$,
find a $U(N)$ gauged non-linear sigma model satisfying
the axioms below.

The low energy action will be determined by a configuration space
$X_N$, a $dN^2$-dimensional manifold  with metric; an action
of $U(N)$ by isometries; and a potential $V$.  The axioms are then

\bl
\item
The classical moduli space,
$$\{X_N|V'=0\}/U(N),$$
is the symmetric
product $\CM^N/S_N$.

\item
The generic unbroken gauge symmetry is $U(1)^N$, while if
two branes coincide the unbroken symmetry is $U(2)\times U(1)^{N-2}$,
and so on.

\item
Given two non-coincident branes at points $p_i\ne p_j$,
all states charged under $U(1)_i\times U(1)_j$ have mass
$m_{ij} = d(p_i,p_j)$.

\item
The action is a single trace (in terms of matrix coordinates),
\eqn\label{eq:singletrace}
S = \Tr (\cdots) .
\enq
\el

The last axiom is familiar in the leading order of open string
perturbation theory (it follows from the definition of Chan-Paton
factors and the disk topology of the world-sheet).  It is also
appropriate for Matrix theory, both so that the action for
block-diagonal matrices will be the sum of that for the individual blocks,
and to get the correct relativistic
dispersion relation for bound states.
It is a non-trivial constraint, as was pointed out by Tseytlin
\cite{Tseytlin}
in the context of the non-abelian Born-Infeld action.

We can give a physical proof that a solution to the problem exists,
in the case that the background is a solution of
the $\ap\rightarrow 0$ limit of  string theory --
just consider D-branes in this background.
Although the stretched strings in this case have masses far above
the string scale, the spacing to the first excited state stays
finite\footnote{
An observation of Steve Shenker.}
and it will still be true that for sufficiently low
energy processes (or length scales $L >> r$) a field theory
description is appropriate.
The axioms can be proven in this context, as the masses of stretched
strings are entirely classical.
On the other hand, $\ap$ corrections could violate the axioms,
in particular axiom 3.

Note that we did not state as an axiom the $U(N)$ gauge action
$X^i \rightarrow U^\dag X^i U$.  We believe that this can be derived,
in the following sense.
When we write an explicit sigma model with matrix coordinates,
we have implicitly chosen a coordinate system for the off-diagonal
components.  For any given coordinate system on $\CM$, the conjecture
is that there exists a choice of matrix coordinates for which the
gauge action will take this form.

To simplify the problem
and get further constraints one can assume additional supersymmetry.
In \cite{dcurve} four real supersymmetries ($\CN=1$, $d=4$) were
assumed, so the target space must be a K\"ahler manifold.
Thus the problem becomes, given a K\"ahler potential $K(z^i,\bar z^\bi)$ on
$\CM$, find a K\"ahler potential $\Tr K_N(Z^i,\bar Z^\bi)$ and
superpotential satisfying the axioms.

We also started with the simplest possible case of one complex dimension,
so the action in this case is a $U(N)$ sigma model with a single matrix
chiral superfield.  By dimensional reduction, a Lagrangian for
D$0$-branes moving in $3+1$ flat and $2$ curved real dimensions can
be obtained.

To give the idea of the analysis, we show how the condition
on the masses of off-diagonal gauge bosons is realized.
Given the free gauge kinetic term $\Re \int d^2\theta W^2$
(in string language, this is constant dilaton), the mass term for the
$ij$ gauge boson is
\eqn\label{eq:gaugemass}
{\p^2 \Tr K\over \p Z^i_{mn} \p Z^{\bar j}_{nm}}
[A,Z^i]_{mn} [A,Z^{\bar j}]_{nm} .
\enq
Expanding around diagonal matrices, one sees
that the second derivative will be a function of the eigenvalues
$z_i \equiv Z_{ii}$:
\eqn
K_N''(z_m, \bar z_m, z_n, \bar z_n) \equiv
{\p^2 \Tr K\over \p Z_{mn} \p Z_{nm}}
\enq
determined by the ordering prescription; for example
\eqa
K_N = Z Z \bar Z \bar Z &\rightarrow& K_N'' = (z_m+z_n)(\bar z_m+\bar z_n)
\nonumber \\
K_N = Z \bar Z Z \bar Z &\rightarrow& K_N'' = 2(z_m \bar z_m+z_n \bar z_n)
\nonumber.
\ena
Computing the commutators in (\ref{eq:gaugemass}) we require
\eqn
K_N'' |z_m-z_n|^2 = d^2(z_m,z_n)
\enq
which we solve for $K''$ and hence for the terms in $K_N$ with up
to two commutators.

At leading order in a normal coordinate expansion,
the result is
\eqn\label{eq:D-kahler}
K_N = \Tr |Z|^2 - {R\over 4} \STr Z^2\bZ^2 + \ldots
\enq
where $\STr$ is the symmetrized trace, normalized as $\STr X^k=\Tr X^k$.

The same ideas can be implemented in ten dimensions.
In the framework of $\CN=1$, $d=4$ gauge theory,
we can describe three complex transverse dimensions
on a general K\"ahler target space, and A. Kato, H. Ooguri and I are in the
process of working out these actions \cite{DKO}.

Rather surprisingly, it appears that the axioms of D-geometry
have no solution in this case unless the manifold $\CM$ is Ricci flat!
The condition that the masses of all strings (with any polarization)
stretched between two D-branes have the same mass
(``the isotropic mass condition'')
 is quite strong and
cannot in general be accomplished with a holomorphic superpotential.

It seems likely that this result is another expression of
the well-known fact that in compactifications of $d=10$ supergravity
to $d=4$, the background will admit $\CN=2$ supersymmetry, allowing
brane solutions with $\CN=1$, only if it is Ricci flat.
\footnote{O. Aharony, S. Kachru and E. Silverstein (unpublished) have made
this observation in the context of brane probe theories, along
with the caveat in the next paragraph.}
Given the relation between these Lagrangians and supermembrane
theory \cite{DHN}, perhaps it can be related to the standard arguments
in this context.
A test of this idea will be to check that, dropping the assumption
of supersymmetry,
the problem can be solved for general metrics.

This argument for Ricci flatness is known to be modified by corrections
to the low energy supergravity Lagrangian (such as the $\ap^3$ term
we mentioned in the superstring-derived Lagrangian) and thus it does
not seem that this proves that D$p$-brane metrics at weak string
coupling must be Ricci flat.
We do see that $\ap$ corrections to the metric must come
with corrections to the isotropic mass condition.

It would be interesting to generalize this to target spaces
of non-trivial topology.  For example, the
question discussed in \cite{DOS} -- what is the
correct Lagrangian for D-branes in an ALE space, reproducing
the supergravity interaction
but remaining non-singular in the orbifold limit? --
should be solvable along these lines.

One would also like to find physical arguments for the higher order
commutator terms.  An interesting example of this is \cite{HashiTaylor}.

\section{Renormalization group in Matrix theory}

The proposal we will now consider is that a gauged sigma model
of the form we just described can be used as a definition of Matrix
theory in curved but non-compact eleven dimensions.
Although we have not completely specified the model,
if it satisfies the axioms, it will reproduce
the leading behavior of the supergravity interaction, at least for
velocity $v$ in the flat directions.

What about the equation of motion ?
Clearly, we must consider the whole family of models for all $N$
(or else take the large $N$ limit) to see how the gauge
theory could know about this.  After all, the model with $N=1$ is perfectly
unitary and consistent with any target space metric, so it does not
know about the supergravity equation of motion.

The story is potentially more interesting for $N \ge 2$ as
quantum corrections can always become large as branes coincide.
In quantum mechanics, these are controlled by $g_s/r^3$.

One concrete proposal for the physical consequences of this is
that since we are seeking models which have a good large $N$ limit
in the sense of \cite{BFSS},
we should try to formulate a large $N$ renormalization group,
whose basic operation is to integrate out a row and column of the
matrix.
Such an approach was first used by Brezin and Zinn-Justin \cite{Brezin}
for the original matrix models of random surfaces,
where it led to good qualitative
results for the critical behavior.
Further motivation for this idea is the relation to string theory.
\cite{DVV,BS}
If we compactify another dimension, the resulting $1+1$ field
theory will undergo the conventional string theory renormalization,
and fixed points will satisfy the equations of motion.
 \cite{Lovelace,CMFP}

We thus start with an action for $N$ D$0$-branes of the type
described in the previous section,
and decompose each matrix into an $(N-1)\times (N-1)$ matrix $X$,
an $N-1$ component vector $v$ and the position of the $N$'th brane
$x_N$:
\eqa\label{eq:msplit}
\left(\matrix{X& v\cr v^\dag& x_N}\right)
\ena

We then integrate out the $N$'th D$0$-brane in two steps.
First, we integrate out the vector $v$.
As long as $x_i \ne x_j\ \forall i\ne j$, this is a completely
well-defined problem.

Let us consider the one loop renormalization of the metric, which
should give a good description for small curvature $R l_{p11}^2 << 1$.
We would like to start with a general target space metric and
from the previous discussion, this will require using
ten-dimensional actions with no supersymmetry assumed.
To illustrate the idea, let us grant that such actions exist with
a kinetic term of the same form as (\ref{eq:D-kahler}) --
in normal coordinates,
\eqn\label{eq:nonsusy}
\Tr (D_t X^i)^2 - {1\over 3} R_{ijkl}(0) \STr DX^i X^j DX^k X^l + \ldots
\enq
In flat ten-dimensional space, the action has maximal supersymmetry,
and zero metric renormalization.
Therefore the leading renormalization of the metric will
come from a one loop diagram with one insertion of the leading
supersymmetry breaking operator, the curvature operator in
(\ref{eq:nonsusy}).
(From (\ref{eq:sigma}), there is also an $R X^6$ term in the potential,
which does not contribute at one loop.)
This leads to the same one-loop diagram as in the usual two-dimensional
sigma model renormalization group, but with a different propagator:
\eqa
&\delta \CL = g^2 \sum_n D_t X^i_{nn} D_t X^j_{nn}
 R_{ikjl}(0) g^{kl}(0) \\
&\qquad\qquad\times \int {dk\over k^2 + |x_n-x_N|^2} + \ldots \nonumber
\ena
The result is no longer UV divergent, and the IR divergence is
controlled for $x_N\ne x_n$.

The second step would be to integrate over $x_N$.
Now the IR divergence at $x_N=x_n$ is not physical, because
the bound state wave function is not singular.
Completing the definition of the RG requires cutting off
the IR divergence, and arguing that physical quantities are independent
of this cutoff.
We do not know enough about the bound states
to make this precise at present, but it is quite plausible that for
small curvature this works the same way as in flat space,
leading to a universal result, on dimensional grounds
\eqn
\delta \CL =  {g^2\over l_{p11}}  R_{ij}(0) \tr D_t X^i D_t X^j
 + \ldots
\enq

Whether we should be able to make sense of flow towards a fixed
point in this framework is not yet clear (it did not have a clear
interpretation in light-cone string theory, either).
What we can say is that the fixed points will be Ricci flat manifolds,
at this order.

\section{The final formulation and locality}

Although we have argued that gauge theory can reproduce the
equation of motion and the leading behavior of the supergravity
interaction, this does not complete our two tests -- we
need to reproduce the exact Green function.
The leading correction in (\ref{eq:green}) vanishes on a Ricci flat
manifold, but the $R^2/d^3$ term does not.

In \cite{DOS} it was shown that this is not possible with a
simple truncation of the weakly coupled open string theory,
or indeed with any model whose expansion to quadratic order
takes the form (\ref{eq:one-loop}).
It is necessary to add additional terms to the Lagrangian,
perhaps higher derivative terms.  However,
we cannot just add the long-range interactions explicitly to the Lagrangian,
as they are singular as $r\rightarrow 0$.

Suppose this were not possible -- what would we conclude ?
Perhaps the simplest way out would be to assert that supergravity is reproduced
only in the large $N$ limit, as in the original conjecture, and
that getting the subleading interactions right requires detailed understanding
of the bound states.

The other out is to assert that
these problems cannot be studied in non-compact backgrounds.
Rather, one must embed the background in a compact background, and even
in the limit that its volume goes to infinity, we need to keep the
additional degrees of freedom.
This cannot be true in the gauge theory definitions of Matrix theory,
where these
states have energies going to infinity in the limit, but perhaps if
sufficiently exotic base theories are used to define Matrix theory this
out will need to be reconsidered.

An interesting related idea is that compactified theories
will be more constrained than the uncompactified theory,
since the base theory is higher dimensional.
Certainly the number of
relevant and marginal perturbations around the Gaussian limit
decreases as we go up in dimension.
This leads to the idea that field theories in
higher dimensions with sensible UV limits are few and far between,
perhaps only coming in a few series, like Lie algebras.

Let us consider the consequences of the modest assumption,
that there exist finitely many such series of field theories with
well-defined UV limits.
Perhaps the simplest is that not all solutions of the supergravity
equations of motion can actually be realized as backgrounds of
the theory!  Let us consider a non-compact space; we would be
saying that only the spaces which are subspaces of the spaces
on our finite list can actually be realized as backgrounds.
There may be some room to extend the list by allowing ``objects''
in the background; as we mentioned earlier it is not clear that one
can make pure deformations of the metric in this way.

This is a highly nonlocal constraint and as such extremely
interesting.  Indeed, such nonlocal effects might shed a new light
on problems such as vacuum selection and
the vanishing of the cosmological constant, which have resisted
real understanding in the context of local physics.

But do we really believe it?
A less radical interpretation is just that any solution of supergravity
is an allowed background, and the 
finite list of theories comes from a finite list of compact manifolds 
which admit such solutions.
On the other hand, there
could be infinitely many theories with non-compact moduli spaces.\footnote{
One might think
that Seiberg's
zero string coupling limit of five-brane
theories \cite{Sei} 
could be used to construct such an infinite set of theories,
by putting the five-branes at points
in a general Ricci-flat 
manifold $\CM_4$ and repeating the construction.
This is not true \cite{SSethi},
essentially because they are bound
(in the language of \cite{BerkoozDouglas},
the defining gauge theory is on the Higgs branch) 
and cannot separate from each other to explore the $\CM_4$ metric.
}

Still, since locality is not at all manifest in Matrix theory, we should
not dismiss such ideas out of hand.
However, I will argue that so far, we have no good reason to believe
in non-locality in Matrix theory
for low energy processes in a time-independent background.
Let us examine the arguments one might make.\footnote{
This section is expanded from the original talk;
I found \cite{Banks} a useful foil.}

One argument is the lack of any manifest locality in the underlying
quantum mechanics.
In particular, the energies of the off-diagonal modes (stretched strings)
are functions of two D-brane positions, leading to apparently
instantaneous interactions between the D-branes.

However, on reflection, we remember that these interactions
are supposed to be one component
of the gravitational interaction, which we know is local.
This type of apparent non-locality is familiar in gauge theories -- for
example, in Coulomb gauge, although there are
explicit non-local interactions in
the Hamiltonian, all non-causal effects are cancelled by other
interactions, leading to a causal theory.
The new element here is that the interactions which we would try
to ``un-gauge fix'' to get a manifestly local formulation are produced
as quantum effects, making it unclear how to realize this locality --
at present.

A better argument is the existence of fundamental extended objects in
the theory.
Does the need to introduce new fundamental objects at each stage of
compactification mean that the theory was non-local?
Yes, in the sense that we don't yet know how to define
the compact theory given the local definition of theory.
But, I would claim, no, not in the usual sense of the word,
that one point can influence another without something propagating
in between.

To define locality
in a theory of extended objects, we must allow for the possibility
that the extended objects themselves have internal degrees of freedom
which can be localized.  Once we accept this, it is not obvious why
having more than one topological class of fundamental object is
essentially
more difficult.\footnote{
One can already make this point in the concrete example of
string field theory.
(See \cite{Martinec,PolSusEtc} for work on locality in this framework.)
In toroidal compactification of superstring theory,
a non-perturbative formulation
must introduce winding strings as new fundamental objects.
The string field $\Psi[X(\sigma)]$
explicitly represents the local degrees of freedom of the
string, and represents the string interaction in terms
of delta functions; schematically
$\int d\sigma\ \delta[X_1(\sigma)-X_2(\sigma)]$, whether or not
we compactify.}

A final motivation is the derived nature of space-time
in Matrix theory -- it arises as the moduli space of a supersymmetric
gauge theory, just as for D-branes at short distances.
Since space-time is not fundamental, we can even imagine situations
with no space-time interpretation -- as Seiberg pointed out in his
talk here,
this may be the general situation when the base theory is strongly coupled.
How can we say that such a theory is local in space-time?

Obviously we don't know how to say it, but that does not mean that
we will never know how to say it -- this is an issue which will take
time to understand.
Again, this issue already arose in perturbative string theory;
non-linear sigma models with highly curved target spaces ($\ap R >> 1$)
are strongly coupled and can be equivalent to
conformal field theory constructions with no obvious
space-time description; nevertheless geometric descriptions have
been found in many cases, as Greene discussed here.

It is hard to prove a no go theorem; in this case the suggestion that
there are compactifications with no definition of locality.
However, superstring duality provides many examples of models with
multiple space-time interpretations.  How could the natural ideas of
locality in the various large volume limits all be valid ?

The simplest proposal one could make for a definition of locality
in this situation is a principle of ``simultaneous locality,''
which asserts the following:

For every large volume limit of the space-time
there exists a definition of locality, which agrees with the
conventional one at large distances but can be extended to cover
the entire parameter space of vacua.  All of these definitions
will be exactly valid in all regimes.

What makes this idea not obviously wrong is that a given space-time
has at most a single large volume interpretation.  All the other
definitions of locality will degenerate and lead to
no constraints at low energies.
However, since they are supposed to be exactly valid,
they will lead to constraints at high energies.

While on this subject, I cannot resist mentioning a striking property:
although nothing in the proposal seems to require it,
many (and perhaps all) of the ``base theories'' used to define Matrix
theory are local.
This locality does not seem to be observable in the large $N$ limit,
but it is certainly visible at finite $N$, and it will be interesting
to interpret this in Susskind's DLCQ proposal.

\section{Conclusions}

We know how to study the behavior of D-branes in curved space at weak
string coupling.
At sub-stringy distances these questions can be reformulated in terms of
the world-volume gauge theory, which is the context which has been
most studied, but this is no more the general prescription than
supergravity was.
A framework valid at all distances has been proposed, but
there are many questions which remain to be
answered in this framework, even very basic ones such
as what metric is seen by
D-branes on various curved spaces.  We have numerous pieces of
evidence that this metric does not always
satisfy the low energy equations of motion of supergravity.

There are even interesting questions about the behavior of D-branes
in curved space in the $\ap\rightarrow 0$ limit, where supergravity
is a good description.  The first question is to write
world-volume theories which reproduce the known physics of enhanced gauge
symmetry.  It turns out that very general consistency conditions
provide much stronger constraints on these actions than one might
have guessed, determining the leading non-abelian terms uniquely
in the simplest case.

\medskip

Matrix theory in curved space is not yet understood.
There is no general proposal for how to do it, and
problems have arisen in the detailed comparison of many conjectures
with supergravity results.
On the other hand, some spectacular successes in flat space motivate
continued efforts to try.

The most straightforward approach is to adapt the non-linear sigma model
approach which we know in string theory.
We proposed here that actions for D-branes in a curved background in
the $\ap\rightarrow 0$ limit are a valid starting point which include
all necessary degrees of freedom in the case of non-compact space
with trivial topology, and we found that such actions can reproduce
the leading behavior of the gravitational interaction as
a one-loop quantum effect.

These actions are only a starting point as there is no reason that
the precise weak coupling string theory action should work in the
large $R_{11}$ limit, but the correct actions will be determined
(perhaps uniquely) by
checking that they precisely reproduce supergravity predictions.
The ability to do this at finite $N$ should provide another
test of Susskind's DLCQ conjecture.

This might serve as a guiding principle for defining the higher
dimensional base theories corresponding to compactifying further
flat directions as well.  For example, we could consider
Matrix theory on $\CM^4\times T^3\times \BR^4$ to get non-linear sigma
models in $3+1$ dimensions.
Although these sigma models appear highly non-renormalizable,
the fact that these spaces are sensible solutions
of supergravity suggests that they exist as sensible field theories
all the way up to energies $E \sim L \rightarrow \infty$ in the
non-compact limit.
Perhaps we will eventually regard their seemingly ill-defined UV behavior as
defined by IR predictions in supergravity.

One can also ask whether taking the large $N$ limit
independently determines the action or leads to new consistency conditions.
We described one framework in which one can study this issue,
a large $N$ renormalization group analogous to the
string world-sheet renormalization group, and
found evidence that its fixed points would have Ricci flat metrics.

The approach we are following
might be regarded as an ``effective theory'' approach and
leaves open the question of what fundamental principles determine
these actions, but the study of these effective theories should provide
valuable information about what these fundamental principles might be.

\medskip
It is a pleasure to acknowledge fruitful collaborations on these topics
with M. Berkooz, D. Finnell, B. Greene, D. Kabat,
A. Kato, M. Li, G. Moore, D. R. Morrison,
H. Ooguri, J. Polchinski, P. Pouliot, S. H. Shenker,
and A. Strominger;
and valuable discussions with P. Aspinwall, C. Bachas,
T. Banks, B. de Wit, D.-E. Diaconescu, M. Green,
S. Kachru, J. Maldacena, J. Schwarz, N. Seiberg, E. Silverstein, L. Susskind,
P. K. Townsend and E. Witten.

This research was supported in part by DOE grant DE-FG02-96ER40959.

\end{document}